# Full-Heusler $Co_2FeSi$ alloy thin films with perpendicular magnetic anisotropy induced by MgO-interface

Yota Takamura[a)], Takahiro Suzuki, Yorinobu Fujino, and Shigeki Nakagawa

*Department of Physical Electronics, Tokyo Institute of Technology, 2-12-1, Ookayama, Meguro-ku, Tokyo 152-8552, Japan*

---

a) Electronic mail: takamura@spin.pe.titech.ac.jp

**ABSTRACT**

The authors demonstrated that $L2_1$-ordered full-Heusler $Co_2FeSi$ (CFS) alloy film with thickness of 100 nm were formed by facing targets sputtering (FTS) method at a substrate temperature $T_S$ = 300 °C. Degrees of $L2_1$- and $B2$- order for the film were 37 %, and 96 %, respectively. Furthermore, full-Heusler CFS alloy thin films with perpendicular magnetic anisotropy (PMA) induced by MgO-interface magnetic anisotropy were successfully formed by the FTS method. The CFS/MgO stacking layers showed PMA when $d_{CFS}$ was 0.6 nm ≤ $d_{CFS}$ ≤ 1.0 nm. The PMA in these structures resulted from the CFS/MgO interfacial perpendicular magnetic anisotropy.

**BODY**

# I. INTRODUCTION

Perpendicular magnetic tunnel junctions (p-MTJs)[1-3] consisting of ferromagnetic electrodes with perpendicular magnetic anisotropy (PMA) have been attracted considerable attention, since they can achieve a low critical current density of current-induced magnetization switching (CIMS). To achieve large tunnel magnetoresistance (TMR) ratio in p-MTJs, highly spin polarized ferromagnetic materials with PMA are required.

Half-metallic ferromagnets (HMFs)[4] are perfectly spin polarized ferromagnetic materials with spin polarization of 100 %. Most of Co-based full-Heusler alloys such as $Co_2FeSi$ (CFS)[5-10], $Co_2FeGe$[6,11-13], $Co_2MnSi$[14] and $Co_2FeSi_{1-x}Al_x$[15], are predicted to be HMFs with higher Curie temperatures than room temperature (RT). Thus, they are considered to be HMF materials appreciate to spintronics devices. Since the half-metallicity of full-Heusler alloys depends on their atomic ordered crystal structures, formation of highly ordered crystal structures are necessary. [16]

Recently, interfacial perpendicular magnetic anisotropy induced by an MgO interface[17] has been realized in full-Heusler $Co_2FeAl$ (CFA) alloy thin films[18], however, full-Heusler CFA alloys are not HMFs according to a band calculation. In the CFA/MgO system, interaction between Fe and O atoms has very important roles in PMA.[19,20] Since half-metallic full-Heusler CFS alloys also include Fe atoms, MgO-induced PMA is expected in the CFS-MgO system.

In this paper, we demonstrated that $L2_1$-ordered full-Heusler CFS alloy thin films were formed by facing targets sputtering (FTS) method. Furthermore, we fabricated a stacking structure consisting of a CFS ultrathin layer and a MgO layer by the FTS method, and observed PMA induced by interfacial magnetic anisotropy between the CFS and MgO layers.

# II. SAMPLE PREPARATION

We fabricated multilayer stacks with CFS/MgO structure on a Pd-buffered layer with an MgO (100) single-crystal substrate. All the samples were prepared with FTS method at a substrate temperature $T_S$ of 300 ºC. A base pressure was $1 \times 10^{-4}$ Pa. Pd buffer layers were sputtered in Kr atmosphere at 0.13 Pa. CFS and Ta layers were sputtered in Ar atmosphere at 0.13 Pa. MgO layers were formed by reactive sputtering of Mg targets in a gas mixture of Ar and $O_2$. Deposition rates for Pd, CFS, MgO, and Ta layers were 0.03 nm/s, 0.05 nm/s, 0.005 nm/s, and 0.06 nm/s, respectively.

## III. RESULTS AND DISCUSSION

Structural properties of 100-nm-thick CFS films formed by FTS were analyzed by means of x-ray diffraction (XRD) with out-of-plane configuration measurements using a Cu $K\alpha$ source. The definitions of the multi-axis goniometer geometries of the out-of-plane configuration are shown in Ref. 12 in detail, where $\phi$ and $\psi$ are the in-plane rotate and the tilted angles, respectively. Figure 1(a) shows a $\theta$–$2\theta$ scan pattern of a CFS(100 nm)/Ta-cap(10 nm) multilayer on a Pd-buffered MgO(100) substrate. CFS(200) and CFS(400) diffraction peaks were clearly observed, indicating that the CFS film was (100)-oriented and had at least the $B2$ structure. Figures 1(b) and 1(c) show $\phi$ scan patterns of the CFS film, where measured at (b) $2\theta = 45.3°$ and $\psi = 45°$, and (c) $2\theta = 27.4°$ and $\psi = 54.7°$. $\phi = 0°$ was defined as a direction of MgO(220). As shown in Fig. 1(b), CFS(220), (202), $(2\bar{2}0)$, and $(20\bar{2})$ diffraction peaks of CFS were clearly observed. These were periodically separated each other with the angle difference of 90°, showing fourfold symmetry in the sample plane. CFS(111), $(1\bar{1}1)$, $(1\bar{1}\bar{1})$, and $(11\bar{1})$ diffraction peaks also exhibited fourfold symmetry (see Fig. 1(c)), which indicated that the CFS film had the $L2_1$ structure. The (220) and (111) diffraction peaks in the $\phi$ scan patterns were separated 45°, which corresponded to the crystal structure of CFS. These results indicate that the CFS film was epitaxially grown.

The order parameters for the 100-nm-thick CFS film was evaluated using the extended Webster model.[7] The degrees of long-range order for the $L2_1$ and $B2$ structure can be evaluated using indices of $S_{L2_1}$ and $S_{B_2}$, respectively. These were calculated from diffraction intensity ratios of CFS(111) to CFS(220) and CFS(200) to CFS(400), which were evaluated by fitting these peaks using the Voigt function. $S_{L2_1}$ and $S_{B2}$ were 37 %, and 96 %, respectively. Note that these values were calculated using the reference intensity ratios for fully $L2_1$-ordered CFS alloys.[21,22] The value of $S_{B2}$ was very high and comparable with that $S_{B2}$ of CFS films formed by rapid thermal annealing at 800 ºC[7]. On the other hand, the $S_{L2_1}$ value was not high enough to obtain the half-metallicity of CFS films. Formation at higher substrate temperatures and post annealing process would be effective. As the discussion above indicated, a $L2_1$-phase CFS film with thickness of 100 nm was successfully formed by the FTS method.

Figure 2(a) shows a schematic stacking structure to obtain MgO-induced magnetic anisotropy in CFS films. CFS/MgO stacking layers were formed on Pd-buffered MgO (100) substrates by FTS method. The CFS/MgO interface is expected to induce PMA when a CFS film is very thin. Figures 2(b)-(e) show magnetic hysteresis loops of the magnetization for the stacks of MgO(100)//Pd(20 nm)/CFS($d_{CFS}$ nm)/MgO(2 nm)/Ta(10 nm) with nominal thickness $d_{CFS}$ = 0.6 nm, 0.8 nm, 1.0 nm, and 2.0 nm. Upper and bottom panels in Figs. 2(b)-(e) show magnetic hysteresis loops for in-plane and out-of-plane magnetic field, respectively. Hysteresis loops for out-of-plane magnetic field with good squareness were observed for the sample of 0.6 nm $\leq d_{CFS} \leq$ 1.0 nm, indicating PMA of the stacking layer. Saturation magnetization $M_S$ and coercivity $H_C$ of the samples with $d_{CFS}$ = 1.0 nm were 800 emu/cc and 710 Oe, respectively. The $M_S$ value was slightly smaller than a bulk value[10] of 1100 emu/cc. (This degradation is discussed in next paragraph.) Unfortunately, magnetic anisotropy energy $K_U$ were not able to be determined, since these in-plane hysteresis loops contained large noises due to the low resolution of the VSM system. The hysteresis loops for the sample for $d_{CFS}$ = 2.0 nm indicated that the magnetic easy axis was in-plane as shown in Fig. 2(e). In this sample, bulk's in-plane magnetic anisotropy energy proportional to thickness (volume) of CFS layers would be larger than PMA energy of CFS/MgO interface.

Origin of the observed PMA was discussed by comparing samples with and without MgO layer. Figure 2(f) shows the hysteresis loops for the sample of 1.0-nm-thick CFS film without MgO layer. This indicated the sample without a MgO layer had in-plane magnetic anisotropy even when CFS layer thickness is 1.0 nm. From these results, we confirmed that the observed PMA resulted from the CFS/MgO interfacial magnetic anisotropy. Note that $M_S$ of the sample without MgO layer were approximately 1000 emu/cc, which was larger than that that of the sample with MgO interface. The reduction of $M_S$ in the sample with MgO may be caused by MgO formation process on CFS, such as interface oxidation of the CFS layer. Further investigation for the crystallinity of these ultrathin CFS films facing MgO layer requires cross-sectional transmission electron microscope (TEM) observation and x-ray photoelectron spectroscopy (XPS).

In summary, we demonstrated that $L2_1$-ordered full-Heusler CFS alloy film with thickness of 100nm were formed by facing targets sputtering (FTS) method at $T_S$ = 300 ℃. Degrees of $L2_1$- and $B2$- order for the film were 37 %, and 96 %, respectively. Furthermore, full-Heusler CFS alloy thin films with perpendicular magnetic anisotropy (PMA) induced by MgO-interface magnetic anisotropy were successfully formed by the FTS method. The CFS/MgO stacking layers showed PMA when $d_{CFS}$ was 0.6 nm ≤ $d_{CFS}$ ≤ 1.0 nm. The PMA in these structure resulted from the CFS/MgO interfacial magnetic anisotropy.

**ACKNOWLEDGEMENT**

A part of this work was supported by JSPS KAKENHI Grant Number 25889021. X-ray diffraction measurements were performed at the Centre for Advanced Materials Analysis (CAMA), Tokyo Institute of Technology, Japan.

## Figure captions

FIG. 1. Out-of-plane XRD patterns of an FTS-formed CFS film for (a) $\theta$–$2\theta$ scan (b) $\phi$ scan for CFS(220) diffraction ($2\theta$ = 45.33°, $\psi$ = 45°), (c) $\phi$ scan for CFS(111) diffraction ($2\theta$ = 27.36°, $\psi$ = 54.7°).

FIG. 2. (a) Schematic of stacking sample structures. In-plane and out-of-plane magnetic hysteresis loops of the stacking layers for (b) $d_{CFS}$ = 0.6 nm, (c) 0.8 nm, (d) 1.0 nm, and (e) 2.0 nm with an MgO layer. (f) 1.0 nm without an MgO layer.

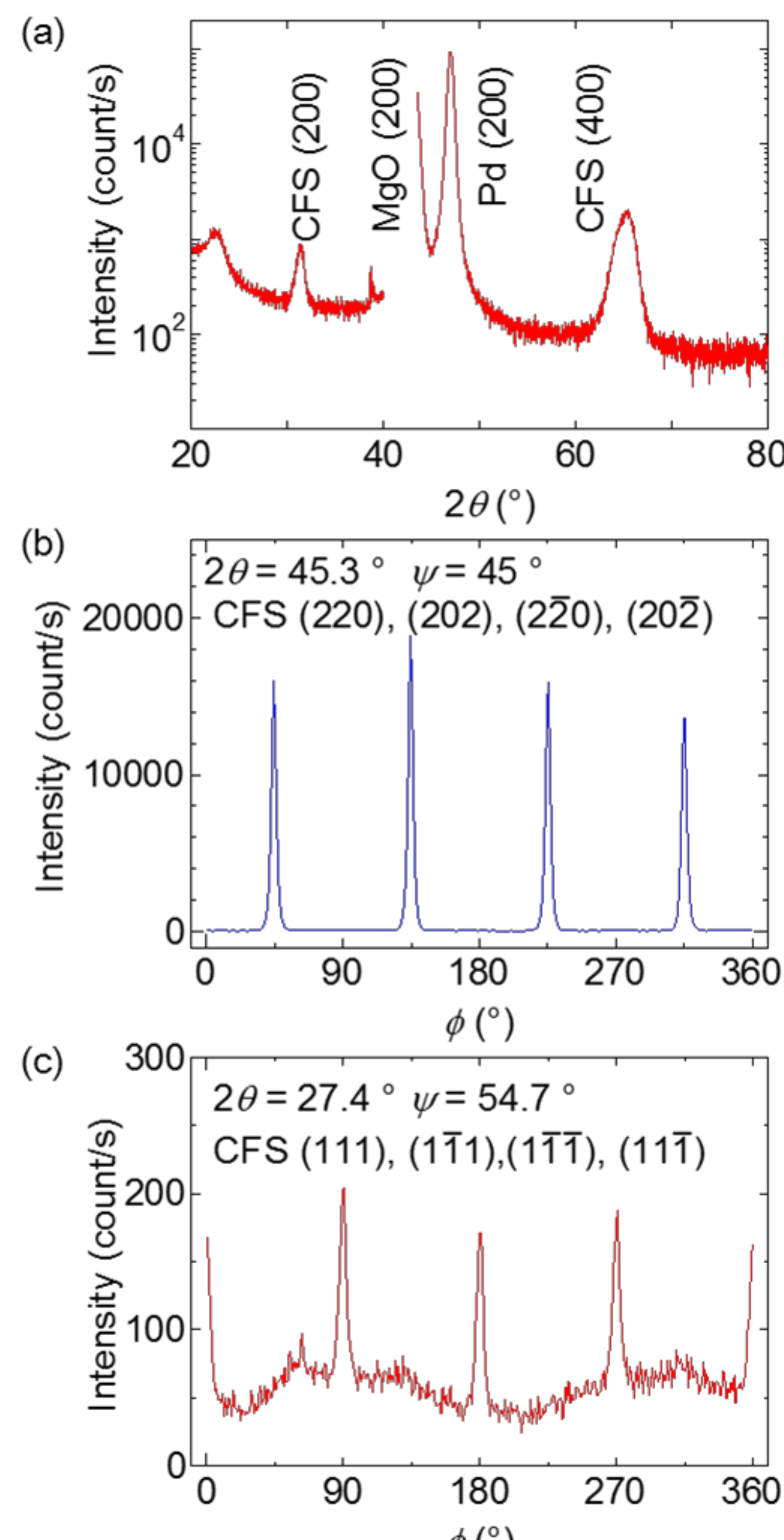

Figure 1 takamura et al.

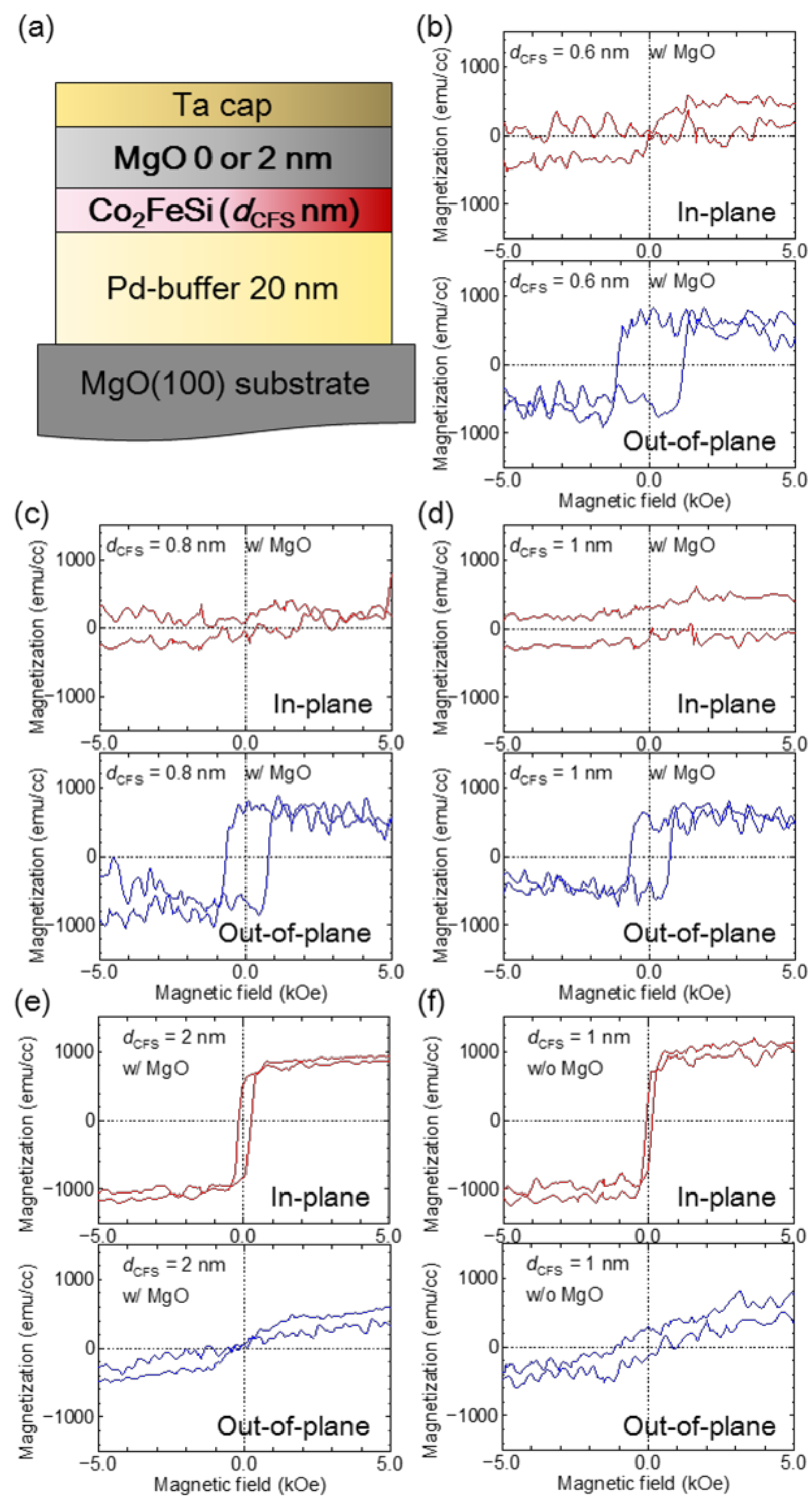

Figure 2 takamura et al.